\documentclass{neuthist18nophoto}

\def\Journal#1#2#3#4{{#1} {\bf #2}, #3 (#4)}

\def\NCA{\em Nuovo Cimento}

\def\PRD{{\em Phys. Rev.} D}
\def\PR{\em Phys. Rev.}

\def\be{\begin{equation}}
\def\ee{\end{equation}}
\def\bea{\begin{eqnarray}}
\def\eea{\end{eqnarray}}

\newcommand{\ba}{\begin{array}{c}}
\newcommand{\baz}{\begin{array}{cc}}
\newcommand{\bad}{\begin{array}{ccc}}
\newcommand{\bav}{\begin{array}{cccc}}
\newcommand{\ea}{\end{array}}

\newcommand{\beq}{\begin{equation}}
\newcommand{\eeq}{\end{equation}}
\newcommand{\meff}{\mbox{$\left| < \! m \! > \right|$}}
\newcommand{\mefff}{\mbox{$\langle m \rangle$}}
\newcommand{\betabeta}{\mbox{$(\beta \beta)_{0 \nu}$}}
\newcommand{\betabetanu}{\mbox{$(\beta \beta)_{2 \nu}$}}

\def\gs{\mathrel{
   \rlap{\raise 0.511ex \hbox{$>$}}{\lower 0.511ex \hbox{$\sim$}}}}
\def\ls{\mathrel{
   \rlap{\raise 0.511ex \hbox{$<$}}{\lower 0.511ex \hbox{$\sim$}}}}
\def\ltap{\ \raisebox{-.4ex}{\rlap{$\sim$}} \raisebox{.4ex}{$<$}\ }
\def\gtap{\ \raisebox{-.4ex}{\rlap{$\sim$}} \raisebox{.4ex}{$>$}\ }
\begin{document}

\vspace*{25mm}
\begin{flushright}
  SISSA 11/2019/FISI \\ 
  IPMU19-0081 
  \end{flushright}

\vspace*{2cm}
\title{The Nature of the Neutrino (Dirac/Majorana) and \\  
Double Beta Decay with or without Neutrinos}

\author{S.T. Petcov~
\footnote{Invited talk at the International Conference 
''History of the Neutrino (1930 - 2018)'', 
September 5 - 7,
2018, Paris, France; published in 
Proceedings of the Conference (eds. M. Cribier, J. Dumarchez and 
D. Vignaud; APC Laboratory, Paris, France, May 2019), p. 417.
}\\
}

\address{SISSA/INFN, Trieste, Italy, and Kavli IPMU, 
University of Tokyo (WPI), Tokyo, Japan   
}

\maketitle

\abstract{
The history of the research associated with the
fundamental problem of the nature of massive neutrinos, 
which can be Dirac particles or Majorana fermions, 
and the related processes of neutrinoless double beta decay, 
$(A,Z)\rightarrow (A,Z+2) + e^- +  e^-$,
and two neutrino double beta decay,
$(A,Z)\rightarrow (A,Z+2) + e^- +  e^- + \bar{\nu_e} +  \bar{\nu_e}$,
is reviewed.
}

\section{Prologue}

 I would like to begin with some general observations.
 
Determining the status of lepton charge conservation and the 
nature - Dirac or Majorana - of massive neutrinos 
is one of the most challenging and pressing 
fundamental problems in present day elementary particle 
physics. The question of the nature of massive 
neutrinos is directly related to the question about the 
underlying basic symmetries of particle interactions.
The massive neutrinos can be Dirac particles if the 
particle interactions conserve some charge
carried by neutrinos, for example, 
the total lepton charge $L = L_e + L_\mu + L_\tau$. 
In this case the neutrinos with definite mass differ from
their respective anti-particles by the value of the 
conserved charge they carry. 
Massive Majorana neutrinos are only possible 
if particle interactions (including the neutrino mass terms) 
do not conserve any lepton charge. 
As is well known, the seesaw and many other 
models of neutrino mass generation 
predict massive neutrinos to be Majorana fermions. 
 
 An inherent problem of the assumption  
of conservation of the total lepton charge $L$ is that 
it is  associated with a global $U(1)_{\rm LC}$ symmetry.
To quote E. Witten \cite{Witten:2017hdv}:
``In modern understanding of particle physics global
symmetries are approximate.''
Similar ideas were expressed by S. Weinberg in 
\cite{SWeinbergCERN1017}.
Thus, the  global $U(1)_{LC}$ symmetry leading to $L=const.$ 
and to massive Dirac neutrinos is expected to be broken, e.g., 
by quantum gravity effects. 
This implies $L$ non-conservation, 
which in turn leads to massive Majorana neutrinos.

The Majorana nature of massive neutrinos manifests itself 
in the existence of processes in which the total lepton charge 
changes by two units, $|\Delta L| = 2$: 
$K^+\rightarrow \pi^- + \, \mu^+ + \,\mu^+$,
$\mu^- + (A,Z)\rightarrow \mu^+ + (A,Z-2)$, etc.
Extensive studies have shown that
the only feasible experiments
having the potential of establishing
that the massive neutrinos are Majorana particles
are at present the experiments searching
for neutrinoless double beta ($(\beta\beta)_{0\nu}-$) decay
\footnote{For an example of a possible alternative
experiment having the potential 
of establishing the Majorana 
nature of massive neutrinos, which, however,
is far from being feasible at present,
see, e.g., \cite{Dinh:2012qb}.}:
\begin{equation}
(A,Z) \rightarrow (A,Z+2) + e^- + e^-\,.
 \label{eq:bb0nu}
\end{equation}
%
Establishing that the total lepton charge 
$L$ is not conserved in particle interactions by observing the 
$\betabeta-$decay would be a 
fundamental discovery (similar to establishing baryon 
number non-conservation (e.g., by observing proton decay)).

 Establishing that the massive neutrinos  $\nu_j$ are Majorana particles 
would be of fundamental importance, as important as the discovery of 
neutrino oscillations, and would have far reaching implications.

 Determining the status of lepton charge conservation and the 
nature - Dirac or Majorana - of massive neutrinos  
are part of a program of research in neutrino physics, which 
aims at shedding light on some of the fundamental aspects 
of neutrino mixing and includes also 
(see, e.g., \cite{Tanabashi:2018oca}):
\\
-- determination of the status of 
CP symmetry in the lepton sector; 
\\
-- determination of the type of neutrino mass spectrum, 
or the ``neutrino mass ordering''; 
\\
-- determination of absolute neutrino mass scale 
or the value of the lightest neutrino mass;\\ 
-- high precision measurement of neutrino mixing parameters. 

\vspace{0.1cm}
The program extends beyond the year 2030. 
Our ultimate goal is to understand at a fundamental level
the mechanism giving rise to neutrino masses and mixing and to
non-conservation of the lepton charges $L_l$, $l=e,\mu,\tau$. 
 The indicated comprehensive experimental 
program of research in neutrino physics 
and the related theoretical efforts are stimulated 
by the fact that the existence of nonzero neutrino masses and 
the smallness of  neutrino masses suggest the existence of new 
fundamental mass scale in particle physics, 
i.e., the existence of New Physics beyond 
that predicted by the Standard Theory.
It is hoped that progress in the theory of
neutrino mixing will also lead, in particular, 
to progress in the theory of flavour and  
to a better understanding of the mechanism of generation 
of the baryon asymmetry of the Universe. 

%
\section{The Early Period 1930 -- 1960} 
%

The history of neutrino physics and double beta decay begin with 
the famous letter of Pauli of December 4, 1930,
in which the hypothesis of existence of neutrinos 
was put forward \cite{Pauli:1930}. 
The next important development was the seminal 
article by Fermi on the $\beta-$decay theory 
\cite{Fermi:1933}. Both were reviewed at this Conference  
by C. Jarlskog \cite{CJ:2018}.

The possibility of two neutrino double 
beta ($\betabetanu-$) decay,
\begin{equation}
(A,Z) \rightarrow (A,Z+2) + e^{-} +  e^{-} +   
\bar{\nu} + \bar{\nu}\,,
\label{eq:bb2nu}
\end{equation}
%
was discussed first in 1935 by 
M.~Goeppert-Mayer in \cite{MGM:1935}.
Goeppert-Mayer realised that some even-even nuclei, 
for which the ordinary $\beta-$decay 
is forbidden, can be still metastable 
since they can undergo a decay 
into a lighter nucleus with emission 
of  two anti-neutrinos and two electrons, 
Eq. (\ref{eq:bb2nu}). She notices that the 
$\betabetanu-$decay can proceed in 2$\rm ^{nd}$ order of Fermi 
$\beta-$decay Hamiltonian $\mathcal{H_{I}^{\beta}}$ 
and thus that the process consists of 
``the simultaneous occurrence of two ($\beta-$decay) transitions, 
each of which does not fulfill the law of conservation 
of energy separately''. Using the Fermi 
Hamiltonian $\mathcal{H_{I}^{\beta}}$ 
she further performs a calculation of 
the rate of the decay (e.g., for a nucleus with $Z=31$) 
and concludes that the $\betabetanu-$decay 
half-life  $T^{2\nu}_{1/2} > 10^{17}$ yr 
(the half-life calculated in \cite{MGM:1935} 
is estimated to be  $T^{2\nu}_{1/2}\sim 10^{20}$ yr). 

In 1936 Gamow and  Teller pointed out \cite{Gamow:1936} that 
the Fermi $\beta-$decay Hamiltonian $\mathcal{H_{I}^{\beta}}$ 
should contain, in addition to the vector 
current $\bar{p}(x)\gamma_{\alpha}n(x)$ postulated by Fermi (describing 
transitions in which the spins and parities of the initial 
and final state nuclei coincide)
also an axial ($\bar{p}(x)\gamma_{\alpha}\gamma_5\,n(x)$)
and/or tensor ($\bar{p}(x)\sigma_{\alpha \beta}\,n(x)$)
currents that could describe the observed $\beta-$decays 
in which the spins of the 
initial and final state nuclei differ by one unit.
The following general four-fermion 
parity conserving $\beta$-decay Hamiltonian
including the Fermi and the Gamow-Teller terms
was proposed in 1936 (see, e.g., \cite{SPVAT36}):
\beq 
\mathcal{H_{I}^{\beta}} = 
\sum_{i=S,P,V,A,T}G_i\,\bar{p}(x)\,O^{i}\,n(x)\,\bar{e}(x)\,O_{i}\,\nu(x) + h.c.,~
O^{i}=I,\gamma_5,\gamma^{\alpha},\gamma^{\alpha}\gamma_5,\sigma^{\alpha \beta}\,.
\label{GT}
\eeq
%
\noindent 
Thus, in addition to the
Fermi vector term (V), the Hamiltonian 
in Eq. (\ref{GT}) includes
scalar (S), pseudo-scalar (P), axial-vector (A) and 
tensor (T) terms ~\footnote{The existence of 
five possible types of four-fermion 
interaction terms (without derivatives) was 
suggested earlier on general grounds, 
i.e., not in connection
with $\beta$-decay theory, in \cite{vonNeum1928}.}.
It is characterised by five constants $G_i$ having
dimension of $(Mass)^{-2}$. 

 In 1937  Majorana published his highly original  
``Teoria Simmetrica  dell'Elettrone e del  Positrone''
\cite{Majorana:1937} 
in which he suggested, in particular,
that neutrinos can be, what we call today, 
``Majorana'' fermions. We are still discussing 
this possibility today. The article by Majorana 
was the subject of the talks by C. Jarlskog and
F. Guerra at this Conference \cite{CJ:2018,G:2018}.    

Soon after the publication of the article 
by Majorana \cite{Majorana:1937},  Racah 
discussed some of the physical implications 
of the Majorana neutrino hypothesis \cite{Racah:1937}. 
He pointed out that if neutrinos are Majorana particles, 
i.e., if they are identical with their antiparticles,
an anti-neutrino produced in $\beta-$ decay 
together with an electron, when interacting 
via the inverse $\beta-$decay reaction with a 
nucleus, can produce an electron again:
\begin{eqnarray}
\label{eq:bdecay}
(A,Z) \rightarrow (A,Z+1) + e^- + {\bar{\nu}}\,,\\
{\bar{\nu}}} + (A,Z+1) \rightarrow (A,Z+2) 
+ e^-\,,~~~{\bar{\nu} \equiv {\nu}\,.
\label{eq:nuMaj}
\end{eqnarray}  
%
For a Dirac neutrino the chain in Eq. (\ref{eq:nuMaj}) is 
impossible: in the inverse $\beta-$decay process 
a Dirac anti-neutrino emitted in $\beta-$decay, Eq. (\ref{eq:bdecay}), 
can produce only a positron 
in the final state. Racah notices that this 
can be used to determine whether neutrinos are 
Dirac or Majorana particles.
Thus, Racah did not discuss
neutrinoless double beta decay, but rather  
pointed out how it might be possible to 
distinguish between Majorana and Dirac neutrinos 
in the processes of inverse $\beta-$decay 
using free neutrino fluxes \cite{BP:1983}.

It was pointed out first by Furry in a well known 
paper from 1939 \cite{Furry:1939} that if neutrinos are 
Majorana particles the process of $\betabeta-$decay, 
Eq. (\ref{eq:bb0nu}), 
%
can take place. In this process two neutrons from the 
initial nucleus transform, by exchanging a virtual Majorana 
neutrino, into two protons of the final state nucleus and two 
free electrons. For Dirac neutrinos the process 
is impossible because the requisite neutrino propagator 
is identically equal to zero~\footnote{This is a consequence, of course, of the total lepton charge 
conservation.}. 
Using the Fermi type four-fermion interaction Hamiltonian
and assuming that the two fermion currents involved are Lorentz 
i) scalars (S), ii) pseudo-scalars (P), 
iii) vectors (V) and iv) axial vectors (A),
Furry performed calculations of the $\betabeta-$decay 
half-life, T$^{0\nu}_{1/2}$, for zero-mass neutrino, $m_{\nu} = 0$.
Comparing his results with the results of 
Goeppert-Mayer \cite{MGM:1935}, Furry concluded 
that due to difference of the phase space factors in 
the two processes, Eqs. ({\ref{eq:bb2nu}) and  ({\ref{eq:bb0nu}),
the $\betabeta-$decay half-life can be shorter by up to 
five orders of magnitude than the  $\betabetanu-$decay half-life:
$T^{0\nu}_{1/2}\sim 10^{-5}\times T^{2\nu}_{1/2}$.

\subsection{The Early Experiments}

  The main motivation for the first experiments searching for 
$\betabetanu$ and $\betabeta$ decays, 
we will call generically ``$2\beta-$decays'',
 was the same as is the motivation 
for the $\betabeta-$decay experiments today~\footnote{A very detailed account of the history of experimental searches 
for $\betabetanu$ and $\betabeta$ decays is given in the review article 
\cite{Barabash:2011}.
} 
-- to answer the 
fundamental question about the nature (Dirac or Majorana) of the neutrino. 
Theory \cite{MGM:1935,Furry:1939} was predicting $\betabeta-$decay half-life 
 $T^{0\nu}_{1/2}\sim 10^{15}$ yr in the case of Majorana neutrino
and  $\betabetanu-$decay half life $T^{2\nu}_{1/2}\sim 10^{20}$ yr 
for Dirac neutrino.

   The first experiment searching for the two processes was performed 
in 1948 by Fireman \cite{Fireman:1948}
with 25 gr of $Sn$ enriched to 54\% with $^{124}Sn$. 
Fireman used Geiger counters for detection of the final state electrons 
and obtained the lower limit $T^{2\beta}_{1/2} > 3\times 10^{15}$ yr 
\footnote{Here and further $T^{2\beta}_{1/2}$ 
should be understood as 
$T^{2\beta}_{1/2} = (\Gamma^{2\nu} + \Gamma^{0\nu})^{-1}\,ln2$, 
where $\Gamma^{2\nu}$ and $\Gamma^{0\nu}$ are the  
$\betabeta$ and $\betabetanu$ decay rates, 
respectively.}.

In the period 1949 -- 1953 positive results 
of searches for  $2\beta-$decays 
were claimed in experiments with 
$^{124}Sn$ \cite{Fireman:1949}, $^{100}Mo$ \cite{Fremlin:1952}, 
$^{96}Zr$  \cite{McCarthy:1953} and 
$^{48}Ca$ \cite{McCarthy:1955}. They were disproved 
by later experiments.

In the same period the first searches for the related processes of 
$(A,Z) \rightarrow (A,Z-2) + e^+ + e^+$ and 
$e^- + (A,Z) \rightarrow (A,Z-2) + e^+$ were performed 
\cite{Fremlin:1952,Berthelot:1953} with negative results.

In these first experiments enriched isotopes 
($^{48}Ca$, $^{94}Zr$, $^{96}Zr$, $^{124}Sn$) were routinely utilised, 
and most advanced (for the time) experimental methods and 
detectors were used (nuclear emulsion, Geiger, proportional, 
and scintillation counters, Wilson chamber).
The detectors were installed relatively deep underground 
(to suppress cosmic ray induced background), and 
active and passive shieldings were employed.  
The sensitivity reached was  $\sim 10^{17}-10^{18}$ yr.
The first experiment in the USSR 
searching for $2\beta-$decays
was performed in 1956 \cite{Dobrokhotov:1956}. 

In 1949,  Inghram and Reynolds performed 
the first geochemical experiment 
searching for $^{130}Te \rightarrow ^{130}Xe + e^{-} + e^{-} 
(+ \bar{\nu} + \bar{\nu})$ 
\cite{Inghram:1949}.
The authors used  ancient ($\sim$several billion years old) minerals
from which Xenon was extracted and subjected to  isotope analysis.
The presence of excess amount of $^{130}$Xe  after accounting for 
various standard contributions from nuclear reactions 
triggered by cosmic rays, etc. would imply that the decay 
$\betabeta$ and/or $\betabetanu$ of $^{130}Te$ took place.
The authors obtained the limit 
$T^{2\beta}_{1/2}(^{130}Te) > 8\times 10^{19}$~yr,
which was  considerably stronger than the limits  
obtained in counter experiments.

 In 1950  Inghram and  Reynolds using the same method 
detected $2\beta^-$ decay of $^{130}Te$ with 
half-life \cite{Inghram:1950}
\begin{equation}
T^{2\beta}_{1/2}(^{130}Te) = 1.4 \times 10^{21}~{\rm yr}\,
~~({\rm Inghram+Reynolds, 1950})\,.
\label{eq:IR1950geoch}
\end{equation}
%
The result of CUORE-0 experiment on $T^{2\nu}_{1/2}(^{130}Te)$ 
obtained in 2016 reads \cite{Alduino:2016}:
\begin{equation}
T^{2\nu}_{1/2}(^{130}Te) = 
[8.2 \pm 0.2 ({\rm stat.}) \pm 0.6 ({\rm syst.})]\times 10^{20}~
{\rm yr}\,. 
\label{eq:CUORE02nu2016}
\end{equation}
%
Thus, apparently the $\betabetanu-$decay of $^{130}Te$ was shown 
to take place first in 
1950 in the geochemical experiment of Inghram and Reynolds.

\subsection{ 1956 - 1960, Related Developments}

In the period 1956 - 1960 important developments took place, 
which had strong impact on the $\betabeta-$decay theory, 
that in turn had significant implications for the searches 
for $\betabeta-$decay.

   In 1956 the neutrino was observed in the inverse $\beta-$decay
experiment of Cowan, Reines et al. \cite{Cowan:1956}. 

  Also in 1956 Lee and Yang put forward the 
(revolutionary) hypothesis of 
non-conservation of parity in $\beta-$decay \cite{Lee:1956}.  
It was confirmed the same year in the experiment of Wu et al.
\cite{Wu:1957}.

 In 1957 the two-component neutrino theory was proposed by  Landau, 
Lee and  Yang, and  Salam \cite{Landau:1957,Lee:1957,Salam:1957}.
Assuming that neutrino mass is zero, $m_{\nu} = 0$, 
Landau, Lee and Yang, and Salam notice that for a massless neutrino 
the left-handed (LH) and right-handed (RH) components of the neutrino field,
$\nu_L(x)$ and $\nu_R(x)$, are independent and postulated 
that only one of them enters the expression of the  
parity non-conserving $\beta-$decay Hamiltonian,  $\mathcal{H_{I}^{\beta}}$. 

In 1958  Goldhaber, Grodzins and Sunyar \cite{Goldhaber:1958} 
have shown that the neutrino emitted in the process 
$e^{-} + ^{152}Eu \rightarrow ^{152}Sm^{*} + \nu 
\rightarrow ^{152}Sm + \nu + \gamma$, has 
a helicity compatible with (-1) and thus it was concluded that the  
neutrino field in $\mathcal{H_{I}^{\beta}}$
is left-handed, $\nu_L(x)$.

 In 1958  Feynman and Gell-Mann \cite{Feynman:1958}, and
 Sudarshan and Marshak \cite{Sudarshan:1958},
proposed the {\it universal $V-A$ current$\times$current 
theory of weak interaction}. These authors postulated 
that all fermion fields in the charged current (CC) weak interaction  
Hamiltonian $\mathcal{H_{I}^{CC}}$, and not only the neutrino fields
\footnote{In addition to $\beta-$decay, $\mathcal{H_{I}^{CC}}$
described also the related processes of $e^\pm$ capture,
$e^-+p\rightarrow n + \nu$, $e^+ + n\rightarrow p + \bar{\nu}$,
inverse $\beta-$decay, $\bar{\nu} + p\rightarrow n + e^+$,
neutrino quasi-elastic scattering  $\nu + n\rightarrow p + e^-$,
and the analogous processes involving $\mu^\pm$ as well as 
$\mu-$decay, $\mu^-\rightarrow e^- + \bar{\nu} + \nu$, 
and $ \bar{\nu} +  e^- \rightarrow \mu^- + \bar{\nu}$.
In \cite{Feynman:1958,Sudarshan:1958} distinction 
between $\nu_e$ and $\nu_{\mu}$ was not made.},
are left-handed, $f(x) \rightarrow f_L(x)$, $f=p,n,\nu,\mu,e$.
They incorporated in $\mathcal{H_{I}^{CC}}$
also the idea of Pontecorvo \cite{Pontecorvo:1947} 
of $\mu - e$ universality of weak interaction. 

Thus,  $\mathcal{H_{I}^{CC}}$ was assumed to have the following elegant form:
\begin{eqnarray}
\label{eq:HwiCC}
& \mathcal{H_{WI}^{CC}} = -\,\mathcal{L_{WI}^{CC}} = 
\frac{G_F}{\sqrt{2}}~j^{\alpha}(x)~(j_{\alpha}(x))^{\dagger}\,,\\
&j_{\alpha}(x)=2\left[\bar{p}_{L}(x) \gamma_{\alpha}n_L(x) + 
\bar{\nu}_{L}(x)\gamma_{\alpha} e_L(x) + 
\bar{\nu}_{L}(x)\gamma_{\alpha} \mu_L(x)\right]\,.
\label{eq:jcc}
\end{eqnarray}
%

 In 1957 Pontecorvo considered the possibility of existence 
of neutrino oscillations \cite{BP:1957,BP:1958}.  
He introduced in \cite{BP:1958} for the first time 
fermion mixing in $\mathcal{H_{WI}^{CC}}$ 
by assuming that the neutrino field $\nu(x)$ in 
$\mathcal{H_{WI}^{CC}}$  
is a linear combination of the fields of two Majorana neutrinos 
$\chi_{1}$ and $\chi_2$ having definite but different masses 
 $m_{1,2} > 0$, $m_1 \neq m_2$, and definite but opposite CP parities,  
$\eta_{1,2CP}$, $\eta_{1CP} = -\,\eta_{2CP}$: 
\begin{equation}
\nu(x) = \frac{\chi_1(x) +\chi_2(x)}{\sqrt{2}}\,.
\label{eq:BP58}
\end{equation} 
%
As it is clear from Eq. (\ref{eq:BP58}), 
Pontecorvo assumed maximal neutrino mixing.

 In 1962 Maki, Nakagawa and Sakata proposed a modification of 
the Nagoya composite model of the proton, neutron and lambda baryons 
with baryon-lepton symmetry \cite{Maki:1962mu}.
They assumed that the neutrinos coupled 
to the electron and the muon in $\mathcal{H_{WI}^{CC}}$, 
Eq. (\ref{eq:jcc}), are different -- a possibility discussed 
since the second half of 1940s  --  and used the notation 
$\nu_e$ and $\nu_\mu$ for them. This was done before 
the results of the two-neutrino Broohaven experiment 
were reported. It was further postulated in \cite{Maki:1962mu} 
that the $\nu_e$ and $\nu_\mu$  fields in 
$\mathcal{H_{WI}^{CC}}$ are linear combinations of the fields 
of two massive Dirac neutrinos $\Psi_{1}$ and $\Psi_2$ having 
different masses, $m^D_{1,2}$,  $m^D_{1} \neq m^D_{2}$:  
\begin{eqnarray}
& \nu_{eL}(x) = \Psi_{1L} \cos\theta_{C} -\, \Psi_{2L}\sin\theta_{C}\,,\\
&\nu_{\mu L}(x) = \Psi_{1L} \sin\theta_{C} + \Psi_{2L}\cos\theta_{C}\,.
\label{eq:MNS62}
\end{eqnarray}
%
By construction (due to the implemented baryon-lepton symmetry)
the mixing angle in (\ref{eq:MNS62}) is what is universally 
called today ``the Cabibbo angle''  
\footnote{In \cite{Maki:1962mu} the authors quote the 1960 article by 
M. Gell-Mann and M. Levy, Nuovo Cim. 16, 605 (1960), 
in which a small mixing parameter 
analogous to the Cabibbo angle was introduced in a footnote.
The article by N. Cabibbo was published in 1963 
in Phys. Rev. Lett. 10, 531 (1963).}.

 In view of the pioneering articles \cite{BP:1958,Maki:1962mu}, 
the neutrino mixing matrix is usually called today the 
Pontecorvo, Maki, Nakagawa and Sakata (PMNS) matrix. 
In the current reference 3-neutrino mixing scheme 
(see, e.g., \cite{Tanabashi:2018oca}) the weak lepton charged 
current has the form:
\begin{equation}
(j^{\rm lep}_{\alpha}(x))^\dagger = 
\sum_{l=e,\mu,\tau} 2\,\bar{l}_{L}(x) \gamma_{\alpha} \nu_{lL}(x)\,,~~~ 
 \nu_{lL}(x) = \sum_{j=1,2,3} U_{lj}\nu_{jL}(x)\,,
\label{eq:3numix}
\end{equation}
%
where  $U\equiv U_{\rm PMNS}$ is the $3\times 3$ unitary PMNS matrix
and $\nu_{jL}(x)$ is the LH component of the field of the neutrino 
$\nu_j$ having mass $m_j$.
%
\section{ The Period 1960 - 1980}
%
%
 The V-A structure of the CC weak interaction 
Hamiltonian (Lagrangian), which was largely confirmed experimentally,
implied that it is possible (and most likely) to have 
$T^{0\nu}_{1/2} >> T^{2\nu}_{1/2}$ instead 
of $T^{0\nu}_{1/2} \sim 10^{-5} T^{2\nu}_{1/2}$.

 Greuling and  Whitten were the first to perform  
a calculation of $\betabeta-$decay rate,  $\Gamma^{0\nu}$,
in the $V-A$ theory assuming that $\nu$ is a Majorana particle
\cite{Greuling:1960}. They had the modern understanding 
of the connection between the lepton charge non-conservation 
and the existence of $\betabeta-$decay. 
These authors showed, in particular, 
that if $\nu$ is a massless Majorana fermion, $m(\nu) = 0$, 
and the electron current in $\mathcal{H_{I}^{CC}}$  has the 
established $V-A$ structure, denoting this current as $(V-A)_e$,
then $\mathcal{H_{I}^{CC}}$  conserves a lepton charge ($L_e$) and 
$\betabeta-$decay is forbidden.  
The $\betabeta-$decay would be allowed 
for Majorana  $\nu$ and  $(V-A)_e$
if $m(\nu) \neq 0$, or if $m(\nu) = 0$, 
but the electron current had a small $V+A$ admixture 
denoted schematically as $(V-A)_e + \eta  (V+A)_e$, where $\eta$
is a free constant parameter.
Greuling and  Whitten showed that for $\eta =0$,
the $\betabeta-$decay rate  $\Gamma^{0\nu} \propto |m(\nu)|^2$, 
while if  $m(\nu) = 0$ but $\eta \neq 0$, 
$\Gamma^{0\nu} \propto |\eta|^2$.

In 1966 Mateosian and Goldhaber
performed a counter experiment with 
11.4 g of $^{48}$Ca (enriched to 96.6\%) \cite{Mateosian:1966}.
In this experiment the scheme 
``detector $\equiv$ source'' was realised for the 
first time. The following limit was obtained:  
$T^{2\beta}_{1/2}(^{48}Ca) > 2\times 10^{20}$ yr and thus the limit of 
$10^{20}$ yr was reached for the first time in a counter experiment.

 In 1967 Fiorini {\it et al.} \cite{Fiorini:1967} made use for the 
first time a Ge(Li) detector in a $\betabeta-$decay experiment. 
The following limit on the 
$\betabeta-$decay half-life of $^{76}$Ge was reported:
$T^{0\nu}_{1/2}(^{76}Ge) > 3\times 10^{20}$ yr.
This limit was improved significantly in 1973 \cite{Fiorini:1973}: 
$T^{0\nu}_{1/2}(^{76}Ge) > 5\times 10^{21}$ yr.
These experiments represented a proof  
of the feasibility of using Ge detectors for high 
sensitivity searches of $\betabeta-$decay. As a consequence,
Ge detectors were used later with increasing sensitivity in a large 
number of experiments (see \cite{Smolnikov} for a complete list and a 
rather detailed description of these experiments). 

 The first experiments employing the ``tracking technique'' 
(visualization of tracks and measurement of the energy of 
two electrons) in the searches for $\betabeta-$decay were performed 
by Wu et al. with $^{48}$Ca and $^{82}$Se using a streamer chamber 
(placed in a magnetic field) and plastic scintillators
\cite{Bardin:1970}. 
The following lower limits were obtained: 
$T^{0\nu}_{1/2}(^{48}Ca) > 2.0\times 10^{21}$~yr 
and $T^{0\nu}_{1/2}(^{82}Se) > 3.1\times 10^{21}$~yr. 

 In the period 1966-1975 several  geochemical 
experiments with $^{130}$Te, $^{82}$Se and 
$^{128}$Te were performed 
\cite{Kirsten:1969,Kirsten:1967,Takaoka:1966,Manuel:1975}.
A major contribution was made by Kirsten et al. 
Kirsten and Muller \cite{Kirsten:1969}
obtained the first compelling evidence 
for the  double beta  decay of $^{82}$Se 
and determined the value of $T^{2\beta}_{1/2}(^{82}Se)$.
These authors found $T^{2\beta}_{1/2}(^{82} Se) = 1.37 \times 10^{20}$ yr 
with an error on the value of  $T^{2\beta}_{1/2}(^{82} Se)$ 
estimated to be about 20\%.
Earlier Kirsten et al. \cite{Kirsten:1967} determined also 
the value of $T^{2\beta}_{1/2}(^{130} Te)$ 
(under conditions which practically did not allow  
any ambiguity in the interpretation of the result obtained).
The following value was reported in \cite{Kirsten:1967}:
$T^{2\beta}_{1/2}(^{130} Te) = 10^{21.34 \pm 0.12}$ yr.
A determination of  $T^{2\beta}_{1/2}(^{130} Te)$ was reported 
also in \cite{Takaoka:1966} where 
the authors found: $T^{2\beta}_{1/2}(^{130} Te) = 
(8.20 \pm 0.64)\times 10^{20}$ yr.
These new determinations of  $T^{2\beta}_{1/2}(^{130}Te)$
confirmed the 1950 result 
obtained in \cite{Inghram:1950}, Eq. (\ref{eq:IR1950geoch}). 
In 1975, Manuel et al. \cite{Manuel:1975}  measured 
$T^{2\beta}_{1/2}(^{130}Te)$ and the ratio 
$T^{2\beta}_{1/2}(^{130}Te)/T^{2\beta}_{1/2}(^{128}Te)$
and used them for the first determination of 
$T^{2\beta}_{1/2}(^{128}Te)$. 
These authors found: 
$T^{2\beta}_{1/2}(^{130}Te)/T^{2\beta}_{1/2}(^{128}Te) 
= (1.59 \pm 0.05)\times 10^{-3}$ 
and  $T^{2\beta}_{1/2}(^{128}Te) = (1.54 \pm 0.17)\times 10^{24}$ yr. 

The quoted geochemically determined values 
of  $T^{2\beta}_{1/2}(^{82} Se)$ and  $T^{2\beta}_{1/2}(^{130} Te)$
are impressively compatible with measurements 
of the $\betabetanu-$decay half-lives of $^{82}$Se and $^{130}$Te 
performed in counter experiments 
much later (see Eq. (\ref{eq:IR1950geoch})
and Table \ref{tab:Tbb2nu}).
%
\section{1960 - 1980: Related Developments}
%

In 1962, following the proposals of Pontecorvo \cite{Pontecorvo:1959sn}
and Schwartz \cite{Schwartz:1960hg},
Danby et al. have demonstrated in an experiment performed at the
Broohaven National Laboratory \cite{Danby:1962nd}
that the neutrinos coupled to the electron
and the muon in the weak charged lepton current are different particles, 
i.e., that $\nu_e \neq \nu_\mu$.  

 In 1969 Gribov and Pontecorvo \cite{Gribov:1969}
considered a Majorana mass term 
for the flavour neutrinos (neutrino fields) 
$\nu_{e}$ and $\nu_{\mu}$ ($\nu_{eL}(x)$ and $\nu_{\mu L}(x)$): 
\begin{equation}
\mathcal{L}_{M}^{\nu}(x) =  
 \,\frac{1}{2}~\nu^{{\rm T}}_{l'L}(x)~C^{-1}~M_{l'l}~\nu_{lL}(x) + h.c.,~~~
l,l'=e,\mu\,,
\label{eq:MajMterm}
\end{equation}
%
where $C$ is the charged conjugation matrix, 
$ C^{-1}\,\gamma_{\alpha}C =-\,\gamma_{\alpha}^{{\rm T}}$.
They  showed, in particular, 
that i) as a consequence 
of $\nu_{l L}(x)$ being fermion fields, the 
mass matrix $M$ in Eq. (\ref{eq:MajMterm}) 
satisfies $M^{\rm T} = M$, and that
ii) the diagonalisation of 
$\mathcal{L}_{M}^{\nu}(x)$ leads to two neutrinos with definite and, 
in general, different masses which are Majorana particles.
\begin{figure}
\centerline{\includegraphics[width=0.6\linewidth]{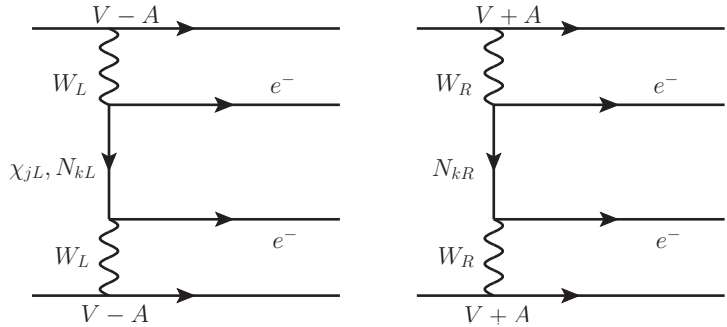}}
\hfill
\caption{Schematic diagrams with exchange of virtual light and 
heavy Majorana neutrinos $\chi_{jL}$ and $N_{kL}$ ($N_{kR}$) 
giving contributions to the $\betabeta-$decay amplitude  
in the case of $V-A$ ($V+A$) CC weak interaction.   
}
\label{fig:LRdiagram}
\end{figure}
%

In 1976 Bilenky and Pontecorvo \cite{Bilenky:1976} 
introduced for the first time Dirac+Majorana neutrino mass term 
which is the most general mass term that can be built from  active LH 
and sterile (singlet) RH neutrino fields $\nu_{\ell L}(x)$ 
and $\nu_{\ell R}(x)$. The  Dirac+Majorana neutrino mass term, 
as is well known, is 
at the basis of the seesaw mechanism 
of neutrino mass generation.
It was shown in \cite{Bilenky:1976} that in the case of 
$n$ LH and $n$ RH neutrino fields 
the diagonalisation of the Dirac+Majorana mass term
leads to $2n$ massive Majorana neutrinos.

The neutrino Majorana and Dirac+Majorana mass terms are integral parts 
of the mechanisms of neutrino mass generations leading to massive Majorana 
neutrinos in gauge theories of electroweak interaction and in GUTs.

  In 1977 Halprin, Primakoff and Rosen \cite{Halprin:1977} 
have performed a calculation of the $\betabeta-$ decay rate 
assuming the presence of an RH admixture in the LH 
weak charged electron current: 
\begin{equation} 
j^{(e)}_{\alpha}(x)= \bar{e}(x)\gamma_{\alpha}\left[(1 - \gamma_5) + 
\eta(1 + \gamma_5)\right]\nu_e(x)\,,
\label{eq:HPR1977a}
\end{equation}
%
where $\eta$ is a constant. The authors of \cite{Halprin:1977}
further assumed that the neutrino $\nu_e$  is a  
``$\gamma_5$ non-invariant Majorana particle'' 
(i.e., that the neutrino field satisfies 
$\nu_e = C(\bar{\nu}_e)^T$). Using the existing experimental 
constraints on the $\betabeta$-decay half-life they 
derived the following limits:
i) $\eta < 5\times 10^{-4}$ for $m(\nu_e) =0$ (``RH mechanism''),
and
ii) $\Gamma^{0\nu}_{th} < 300\,\Gamma^{0\nu}_{exp.lim}$ 
 for  $\eta$=0 and $m(\nu_e) = 60$ eV (the upper limit 
on $m(\nu_e)$ at the time), 
$\Gamma^{0\nu}_{th}$ and $\Gamma^{0\nu}_{exp.lim}$
being the theoretically predicted and 
the maximal experimentally allowed decay rates.
In \cite{Halprin:1977} the first $\betabeta-$decay 
bounds were obtained on the mass $M_N$ of a heavy 
Majorana neutrino $N_e$ coupled to the electron in 
the weak charged lepton current 
(see Fig. \ref{fig:LRdiagram} where $\chi_{jL} \equiv \nu_{eL}$ 
and $N_{kR(L)} \equiv N_{eR(L)}$):
\begin{equation} 
j^{(e)}_{\alpha}(x)= \bar{e}(x)\,\gamma_{\alpha}(1 - \gamma_5)\,\nu_e(x) 
+ U_{eN}\,\bar{e}(x)\,\gamma_{\alpha}(1 \pm \gamma_5)\,N_e(x)\,.
\label{eq:HPR1977b}
\end{equation}
%
Using the existing data on $\betabeta$-decay 
the following bounds were reported for 
$U_{eN} = 1$  \cite{Halprin:1977}: $M_N < 1$ keV, or $M_N > 3$ GeV. 
The current combined $|U_{eN}|^2-M_N$ limits are shown 
in Fig. \ref{fig:UeNvsMN}.
\begin{figure}
\centerline{\includegraphics[width=0.8\linewidth]{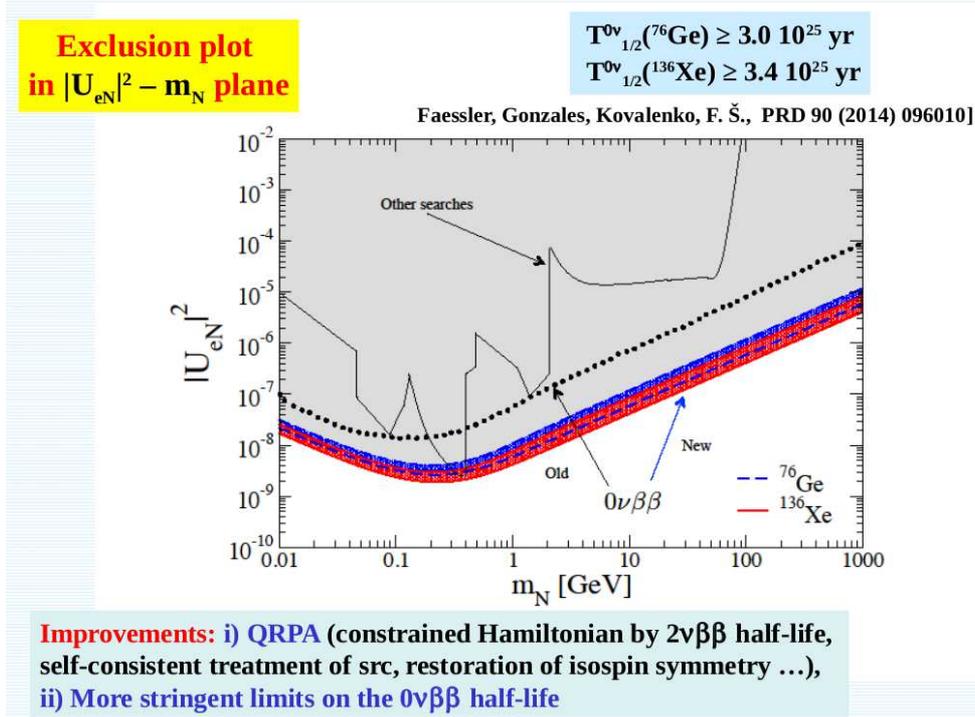}}
\caption{The exclusion plot in the $|U_{eN}|^2 - M_n$ plane 
 ($m_N \equiv M_N$) from the limits on the half-lives of 
 $^{76}$Ge and $^{136}$Xe indicated in the figure. 
(Figure from \protect\cite{Faessler:2014} 
 updated by F. \v Simkovic in 2017.)
}
\label{fig:UeNvsMN}
\end{figure}
%

  Starting from 1976 gauge theories 
(extensions of the Standard Model, GUTs)
naturally incorporating $L$-non-conservation 
and massive Majorana neutrinos were proposed 
(see \cite{HTM}, for example). 
In particular, the enormous disparity between the 
magnitudes of the neutrino masses and the masses 
of the charged leptons and quarks 
found a natural explanation by the 
seesaw mechanism \cite{Seesaw} and the Weinberg dimension 
5 effective operator \cite{Weinberg1979}, both of which 
implied $L-$non-conservation and 
massive Majorana neutrinos.

\section{The Period 1980 - 1990: Theoretical Activity}

 In this period there was a renewed interest in the process of 
$\betabeta-$decay (see further). It was stimulated, in particular,
by the development of gauge theories of electroweak interactions 
and GUTs in which, as we have indicated, massive Majorana neutrinos 
appear naturally. In addition, the leptogenesis scenario 
of the origin of baryon asymmetry of the Universe, 
relating  the neutrino mass generation 
via the  seesaw mechanism (leading to Majorana neutrinos)  
with the generation of the matter-antimatter 
asymmetry of the Universe, was proposed 
\cite{FY1986}. A massive neutrino 
(with a mass of $\sim$ 40 eV) was considered in that period 
as a plausible (hot) dark matter candidate.
And Lubimov et al. claimed in 1980 the observation
of a neutrino mass of $\sim$ 20 eV in an ITEP 
tritium $\beta$-decay experiment 
\footnote{This claim was proven to be incorrect much later by 
independent experiments (see the contribution of  M. Goodman
in these Proceedings \cite{MGoodman}).}.

We will discuss  next some relevant theoretical developments 
which took place in the period 1980 - 1990.
 
 \subsection{The Majorana Phases}

 In 1980 in a JINR (Dubna) preprint published in May, 
Bilenky et al. \cite{Bilenky:1980} 
pointed out that if the massive neutrinos are 
Majorana particles, the neutrino mixing matrix 
contains additional physical CP violation (CPV) 
``Majorana'' phases
in comparison to the number of ``Dirac'' phases present in the 
case of massive Dirac neutrinos.
The authors of \cite{Bilenky:1980} showed, 
in particular, that in the case of $n$ lepton families, the 
$n\times n$ unitary neutrino mixing matrix $U_{\rm PMNS}$ 
contains, in general, 
$(n-1)$ Majorana CPV phases in addition to the 
$(n-1)(n-2)/2$ Dirac CPV phases, or 
altogether $n(n-1)/2$ CPV phases. 
It was noticed in \cite{Bilenky:1980} (see also \cite{Bilenky:1984}) that 
in this case the PMNS matrix can be cast in the form: 
\begin{eqnarray}
&U_{\rm PMNS} = VP\,,~V:~n\times n~{\rm unitary,~includes}~(n-1)(n-2)/2 
~{\rm Dirac~CPV}\,,\\
&P = {\rm diag}(1,e^{i\frac{\alpha_{21}}{2}},e^{i\frac{\alpha_{31}}{2}},...,
e^{i\frac{\alpha_{n1}}{2}})\,:~\alpha_{k1},~k=2,...,n~-~(n-1)~
{\rm Majorana~CPV}\,.  
\label{eq:CPVDM}
\end{eqnarray}
%
In the reference case of 3 families, these results imply that  
$U_{\rm PMNS}$ contains one Dirac phase (in $V$) 
and two Majorana CPV phases (in $P$).
Even in the mixing involving only 2
massive Majorana neutrinos there
is one physical CPV Majorana
phase. In contrast, the CC weak interaction
is automatically CP-invariant in the case of
mixing of two massive Dirac neutrinos
or of two quarks.

Similar conclusions concerning the existence of Majorana CPV phases were 
reached in~\cite{Schechter:1980} and~\cite{Doi:1980}, which appeared in preprint forms respectively 
in June and July of 1980.

In \cite{Bilenky:1980} it was proven also that the 
flavour neutrino and antineutrino vacuum oscillation 
probabilities, $P(\nu_l\rightarrow\nu_{l'})$ and 
$P(\bar{\nu}_l\rightarrow\bar{\nu}_{l'})$, are not 
sensitive to the Majorana phases. It was shown in 
\cite{Langacker:1987} that the same result holds 
also when the flavour neutrino and antineutrino 
oscillations
take place in matter. These results implied that it is impossible 
to get information about the nature of massive neutrinos 
in experiments studying $\nu_l\rightarrow\nu_{l'}$ 
and/or $\bar{\nu}_l\rightarrow\bar{\nu}_{l'}$
oscillations.

 The authors of~\cite{Doi:1980} realised that 
in the case when the $\betabeta-$decay is 
induced by exchange of light Majorana neutrinos 
$\chi_j$ (with masses $m_j \ltap 1$ MeV, see Fig. \ref{fig:LRdiagram}), 
the $\betabeta-$decay rate 
$\Gamma^{0\nu}\propto |\mefff|^2$, 
where $\mefff$ is the $\betabeta-$decay 
effective Majorana mass parameter, 
\begin{equation}
\meff = |\sum_{j=1,2,3}m_j\,U^2_{ej}| =
\left | m_1 |U_{e 1}|^2 
+ m_2 |U_{e 2}|^2~e^{i\alpha_{21}}
 + m_3 |U_{e 3}|^2~e^{i(\alpha_{31}-2\delta)} \right|\,,
\label{eq:meff}
\end{equation}
%
and we have given the expression of $\meff$ 
for 3-neutrino mixing and the standard parametrisation 
of the PMNS matrix  \cite{Tanabashi:2018oca}. 
In Eq. (\ref{eq:meff}) $\delta$ is the Dirac CPV phase.
It was also noticed in \cite{Doi:1980} that because of the 
presence of the Majorana phases 
there is a possibility of 
cancellation between the different terms 
in $\meff$, and thus of a strong suppression 
of $\Gamma^{0\nu}$.

In 1981 Wolfenstein derived the CP invariance constraints 
on massive Majorana neutrinos and on the Majorana CPV phases
\cite{Wolfenstein:1981} (see also \cite{Bilenky:1984,Kayser:1984}).
He showed that in the case of CP invariance, 
the Majorana neutrinos with 
definite mass $\chi_j$ have definite CP parities 
equal to $\pm i$:
\begin{equation}
U_{CP}\,\chi_j(x)\,U^{\dagger}_{CP} = \eta^{CP}_j\,\gamma_{0}\,\chi_j(x_p)\,,~~~ 
\eta^{CP}_j = \pm i\,,
\label{eq:MajnuCPpar}
\end{equation}
%
where $\chi_j(x)$ is the field of $\chi_j$ and 
$U_{CP}$ is the unitary operator of CP transformation.
The Majorana phases $\alpha_{j1}$, 
$j=2,3,...$, take the CP conserving values $k_{j1}\pi$, 
$k_{j1}=0,1,2,...$. Correspondingly, the phase factors 
in the expression for $\meff$ take values 
$e^{i\alpha_{j1}} = \pm 1$, which represent the relative 
CP parities of the Majorana neutrinos 
$\chi_j$ and  $\chi_1$.

 It follows from  
Eq. (\ref{eq:meff}) and the preceding discussion, in particular, 
that the maximal and minimal values 
of $\meff$ are obtained in the case of CP invariance and 
are determined by the CP conserving 
values of the Majorana phases 
\footnote{The CP conserving values of 
$\delta = 0,\pi$ do not have an effect on 
$\meff$, Eq. (\ref{eq:meff}).} 
which thus play 
very important role in the phenomenology of 
$\betabeta-$decay. 

\subsection{The Process with Majoron Emission}

In 1981 it was pointed out that the spontaneous breaking of 
the global $U(1)_{\rm LC}$ symmetry, associated with 
the $L$ conservation, leads
(in accordance with the Goldstone theorem
\cite{Goldstone:1961eq})
to the presence in the theory 
of a massless scalar particle, $\phi^{0}$, called ``Majoron'' 
\cite{Majoron}. In this case the process of neutrinoless 
double beta decay with emission of a Majoron becomes possible:
\begin{equation} 
(A, Z) \rightarrow (A, Z + 2) + e^{-} +  e^{-} + \phi^{0}\,.
\label{eq:bbMaj}
\end{equation}
%
The searches for this process continue today.

\subsection{The ``Black Box'' Theorem}

In 1982 Schechter and Valle published the 
``Black Box'' theorem \cite{Schechter:1982}
 which stated that the observation of  $\betabeta-$decay 
implies (under very general conditions)
the existence of a Majorana mass term of $\nu_e$ and thus 
massive Majorana neutrinos independently of the mechanism 
which triggers the decay (Fig. \ref{fig:BBox1}).
\begin{figure}
\centerline{\includegraphics[width=0.8\linewidth]{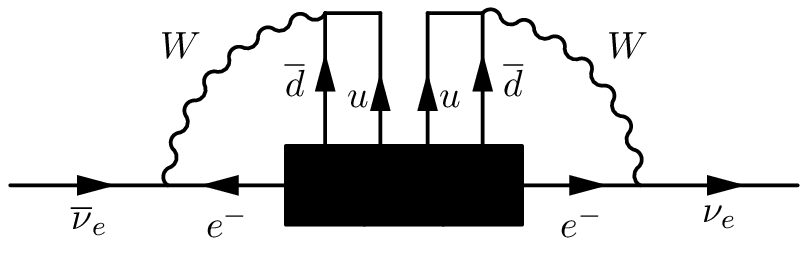}}
\caption{The diagram generating Majorana mass term of $\nu_e$ 
assuming the existence of $\betabeta-$decay.
}
\label{fig:BBox1}
\end{figure}
%
This theorem became serious theoretical motivation for 
continuing the experimental searches for  $\betabeta-$decay 
with increasing sensitivity.

 In 2011 a study of the  magnitude of the 
Majorana mass term of $\nu_e$ generated 
according to the ``Black Box'' theorem 
was performed in \cite{Duerr:2011}. The Black Box operators induce Majorana 
neutrino mass term at four-loop level (see Fig. \ref{fig:BBox1}).
This warrants the statement that an observation of neutrinoless double 
beta decay guarantees the Majorana nature of neutrinos.
The authors of \cite{Duerr:2011} evaluated the Schechter-Valle (Black Box) 
theorem quantitatively by considering the most general Lorentz 
invariant Lagrangian consisting of point-like 
operators for neutrinoless double beta decay 
and calculating the corresponding 
radiatively generated neutrino masses.
They found that these neutrino masses ``are many orders 
of magnitude smaller than the observed neutrino 
masses and splittings''. 
As an example they give the following prediction 
of the value of the Majorana neutrino mass generated 
by the Black Box diagram (see \cite{Duerr:2011} for details of the 
relevant calculation): $\delta m_\nu= 5\times 10^{-28}$ eV.
The authors of \cite{Duerr:2011} thus concluded:
``Although the principal statement of the Schechter-Valle 
theorem remains valid, we conclude that the Black Box diagram 
itself generates radiatively only mass terms which are many orders 
of magnitude too small to explain neutrino masses. 
Therefore, other operators must give the leading 
contributions to neutrino masses, which could be of Dirac or 
Majorana nature.''

\subsection{Nuclear Matrix Element Calculations}

In the period 1984 -- 1986 the first more detailed 
calculations of the nuclear matrix elements (NME) of 
$\betabeta$ and $\betabetanu$ decays were performed.

In \cite{Haxton:1984} it was pointed out, in particular,
that the NME of $\betabeta$ and $\betabetanu$ decays 
are very different in structure and are not related.
The NME were obtained within the shell model by considering 
the weak coupling approximation scheme
\footnote{The NME obtained in  \cite{Haxton:1984}
are significantly smaller (up to factor 2-3) when
compared with nuclear shell model matrix elements
calculated recently with larger model space, more advanced
nucleon-nucleon interaction
and by avoiding the previous many-body simplifications.}.

  The 1985 article by Doi et al. \cite{Doi:1985}
represents a basic theoretical work on the calculation of the NME 
for the $(\betabetanu$ and $\betabeta$ decays.
The energy and angular distributions of the two e$^-$ were  
calculated for the ``standard'' light Majorana neutrino exchange 
and the RH current mechanisms. They authors 
pointed out to the possibility to distinguish experimentally 
between the two mechanisms using the difference between 
the respective distributions.

 In 1986 in was shown in \cite{Vogel:1986}
that the inclusion of the 
particle-particle interaction in a nucleus, 
characterised by a constant $g_{pp}$, within the 
quasi-particle random phase approximation (QRPA) method  
(which was usually disregarded
in the QRPA calculations of the ordinary beta decay matrix elements
- only particle-hole interaction was considered)
permits to calculate the NME and the rate of the  
$\betabetanu-$decay with relatively 
small uncertainties and to reproduce the measured 
$\betabetanu-$decay half-lives 
\footnote{
The authors found also that the results obtained 
exhibit very strong dependence on the constant  $g_{pp}$.
}.
This result led to an extensive 
use of the QRPA models for the calculation of NMEs for
$\betabetanu$ and $\betabeta$ decays.

\section{Experimental Activity: 1980-2000}  

 The experimental activity related to $2\beta-$decays 
increased significantly in the period 1980--2000. 
The use of detectors with passive and 
active shielding located deep underground,
built with low background materials
led to a substantial reduction of background. 
Large (several to few$\times$10 kg) relatively cheap high purity germanium 
detectors became available and 
several experiments were performed with $^{76}$Ge.
These developments led to an increase of sensitivity 
by several orders of magnitude: several collaborations 
reported limits on T$^{0\nu}_{1/2}$ of $\sim10^{23}$-$10^{25}$ yr.

The limit $T^{0\nu}_{1/2}(^{76}Ge) > 1.2\times 10^{24}$ yr (90\% C.L.)
was reported by D. Caldwell \cite{Caldwell:1991} 
using 7.2-kg high-purity 
natural Ge (7.8\% $^{76}$Ge) semiconductor 
detectors (inside of NaI anticoincedence shield, which was 
inside a high purity Pb shield, 
at a depth of 600 m w.e.)

In 1987 the ITEP/YePI collaboration used for the first time 
semiconductor Ge(Li) detectors
grown from Ge enriched in $^{76}$Ge at 85\% to search for 
$\betabeta$, $\betabetanu$ and Majoron decays \cite{Vasenko:1987}.
The following results were obtained: 
$T^{0\nu}_{1/2}(^{76}Ge) > 2\times 10^{24}$ yr (68\% C.L.),
$T^{2\nu}_{1/2}(^{76}Ge)=(9 \pm 1)\times 10^{20}$ yr,
$T^{0\nu,M}_{1/2}(^{76}Ge) > 1.2 \times 10^{20}$ yr, 
$T^{0\nu,M}_{1/2}(^{76}Ge)$ being the half-life of 
the decay with Majoron emission, Eq. (\ref{eq:bbMaj}).
This progress paved the way for 
 the Heidelberg--Moscow \cite{HM:2001} 
and IGEX \cite{Aalseth:2002} experiments with enriched $^{76}$Ge,  
which reached record sensitivity to 
T$^{0\nu}_{1/2}$($^{76}$Ge) of $\sim 10^{25}$ yr. 

 The limit $T^{0\nu}_{1/2}(^{136}Xe) > 3.4\times 10^{23}$ y was 
obtained \cite{Luescher:1998} in a time projection chamber experiment 
with 3.3-kg Xe enriched in $^{136}$Xe to 62\%, 
in which the sum of the energies and the individual energies 
of the two electrons were measured, events with the
simultaneous emission of 2e$^-$  from one point were
selected and the tracks of electrons were reconstructed.

 In 1984 Fiorini and Niinikoski discussed the possibility 
(following the idea of G.V. Mizelmaher, B.S. Neganov, V.N. Trofimov 
from JINR Dubna (Communication JINR P8-82-549, Dubna, 1982 (in Russian))
to use low-temperature (bolometer) detectors to search for 
$(\beta\beta)_{2\nu}$ and $\betabeta$ decays \cite{Fiorini:1984}. 
The proposed method was successfully realised by the 
Milano group in CUORICINO and  CUORE experiments.

 A breakthrough experimental result was reported 
in 1987 by  Moe et al. \cite{Moe:1987}: for the first 
time the  $\betabetanu-$decay was  observed 
in a laboratory experiment. 
36 events of  $\betabetanu-$decay of $^{82}$Se were detected 
in a direct counter experiment with a time projection chamber.  
The following half-life was measured:
$T^{2\nu}_{1/2}(^{82}Se)=1.1_{-0.3}^{+0.8} \times 10^{20}$ yr.

 In the 1990s the  $\betabetanu-$decay 
was observed and its half-life measured  of 
i) $^{76}$Ge \cite{Vasenko:1987}, 
ii) $^{100}$Mo, $^{150}$Nd and $^{48}$Ca \cite{Moe:1991}, 
iii) $^{100}$Mo and $^{116}$Cd \cite{Ejiri:1991},
iv) $^{100}$Mo, $^{116}$Cd, $^{82}$Se  
and $^{96}$Zr by NEMO--2 experiment \cite{NEMO2:1995} 
(in which the energy spectra and angular distributions 
of electrons were also measured).

 In the discussed period a number of 
geochemical experiments were performed as well.  
In 1993 the first geochemical experiment with $^{96}$Zr 
was realised and the following  half-life for 
the $^{96}$Zr-- $^{96}$Mo transition was obtained \cite{Kawashima:1993}: 
$T^{2\beta}_{1/2}(^{96}Zr-^{96}Mo)=(3.9 \pm 0.9)\times 10^{19}$ yr.
The values of $T^{2\beta}_{1/2}(^{82}Se)$, 
$T^{2\beta}_{1/2}(^{130}Te)$ and $T^{2\beta}_{1/2}(^{128}Te)$ were 
again discussed in the period of interest. 
Kirsten et al. \cite{Kirsten:1986} reported 
$T^{2\beta}_{1/2}(^{82}Se) = (1.30 \pm 0.50)\times 10^{20}$ yr,
$T^{2\beta}_{1/2}(^{130}Te) = (1.63 \pm 0.14)\times 10^{21}$ yr,
and stated that in what concerns $T^{2\beta}_{1/2}(^{130}Te)$,
``the majority of data from all sources is $ >  1.5\times 10^{21}$ yr.''
At the same time Manuel reported \cite{Manuel:1986}: 
$T^{2\beta}_{1/2}(^{130} Te) = (7 \pm 2)\times 10^{20}$ yr, 
$T^{2\beta}_{1/2}(^{82}Se) = 1\times 10^{20}$ yr 
and  $T^{2\beta}_{1/2}(^{128}Te)/T^{2\beta}_{1/2}(^{130} Te) 
= 2\times 10^{3}$.
The inverse of the last ratio was determined with a 
high precision by  Bernatowicz et al. \cite{Bernatowicz:1993}:
$T^{2\beta}_{1/2}(^{130}Te)/T^{2\beta}_{1/2}(^{128} Te) 
= (3.52 \pm 0.11)\times 10^{-4}$. 
Takaoka et al. \cite{Takaoka:1996} used this result together 
with $T^{2\beta}_{1/2}(^{130} Te) = (7.9 \pm 1.0)\times 10^{20}$ yr 
obtained in their experiment 
to determine the half-life of 
$^{128}$Te: $T^{2\beta}_{1/2}(^{128}Te)= (2.2 \pm 0.3)\times 10^{24}$ yr.
Bernatowicz et al. \cite{Bernatowicz:1993} reported also:
$T^{2\beta}_{1/2}(^{130}Te) = (2.7 \pm 0.1)\times 10^{21}$ yr 
and  $T^{2\beta}_{1/2}(^{128}Te) = (7.7 \pm 0.4)\times 10^{24}$ yr.
The origins of the discrepancies between the results on 
$T^{2\beta}_{1/2}(^{130}Te)$ and $T^{2\beta}_{1/2}(^{128}Te)$ 
obtained by the different authors were not understood at the 
time; later the smaller values were proven to be correct.

 Let us add that in the 1980s compelling experimental 
evidences have been obtained, which showed that the 
neutrino $\nu_{\tau}$ coupled to 
the $\tau$ lepton in the weak charged lepton current 
is of a new type (see \cite{Blondel:2018}), i.e., that  
$\nu_{\tau}\neq \nu_e,\nu_\mu,\bar{\nu}_e,\bar{\nu}_\mu$.  
In 2000 the DONUT experiment had directly observed 
the $\nu_{\tau}$ interaction with matter in which the $\tau$ 
lepton was produced \cite{DONUT:2000}.

%
\section{Developments in the Period 2000 -- 2010}
%
%
  In 1998 the Super Kamiokande experiment provided 
compelling evidence for oscillations of the  
atmospheric $\nu_\mu$ and $\bar{\nu}_\mu$ 
caused by non-zero neutrino masses and 
neutrino mixing \cite{Fukuda:1998mi}, 
which was further elaborated in \cite{SKdip04}. 
In 2001 -- 2002 flavour conversion of the solar 
$\nu_e$ was proven to take place \cite{ahmad01}.
These remarkable discoveries, for which
Takaaki Kajita 
(from the Super-Kamiokande Collaboration)
 and Arthur McDonald 
(from the SNO Collaboration)
 were awarded the Nobel Prize for Physics in 2015,
had far reaching implications in particle physics in general, and 
for the searches for $\betabeta-$decay in particular.
Further, disappearance of reactor $\bar{\nu}_e$
due to oscillations has been observed
in the KamLAND experiment \cite{KL162},
while strong (to compelling) evidence for $\nu_{\mu}$ disappearance
due to oscillations were obtained also in the long-baseline
accelerator neutrino experiments K2K \cite{ahn06}
and  MINOS \cite{MINOS-initial}.

In the reference 3-neutrino mixing framework 
and in the standard parametrisation of the PMNS matrix 
(see, e.g., \cite{Tanabashi:2018oca}) 
the oscillations of the atmospheric and accelerator
$\nu_\mu$ and $\bar{\nu}_\mu$  and the flavour conversion of
the solar $\nu_e$ to a good approximation 
are driven respectively 
by the parameters $\Delta m^2_{31}$, $\theta_{23}$ and 
$\Delta m^2_{21}$, $\theta_{12}$,
where $\Delta m^2_{ij} \equiv m^2_i - m^2_j$, $m_i$, $i=1,2,3$,
being the mass of the neutrino $\nu_i$,  
$\sin^2\theta_{23} \equiv |U_{\mu 3}|^2/(1 - |U_{e 3}|^2)$ 
and $\sin^2\theta_{12} \equiv |U_{e2 }|^2/(1 - |U_{e 3}|^2)$. 
Here $U_{lj}$, $l=e,\mu,\tau$, $j=1,2,3$, are elements of the 
PMNS matrix. 
In 1999 the CHOOZ experiment obtained a strong limit on 
the third mixing angle in the 3-neutrino mixing matrix,
$\theta_{13}$  \cite{CHOOZ1999}: $\sin^22\theta_{13} < 0.10$ (90\% CL) at
$|\Delta m^2_{31}|= 2.5\times 10^{-3}~{\rm eV^2}$,
where $\sin^2\theta_{13} = |U_{e 3}|^2$ (Fig. \ref{fig:CHOOZ}).
\begin{figure}
\includegraphics[width=6cm,height=8cm]{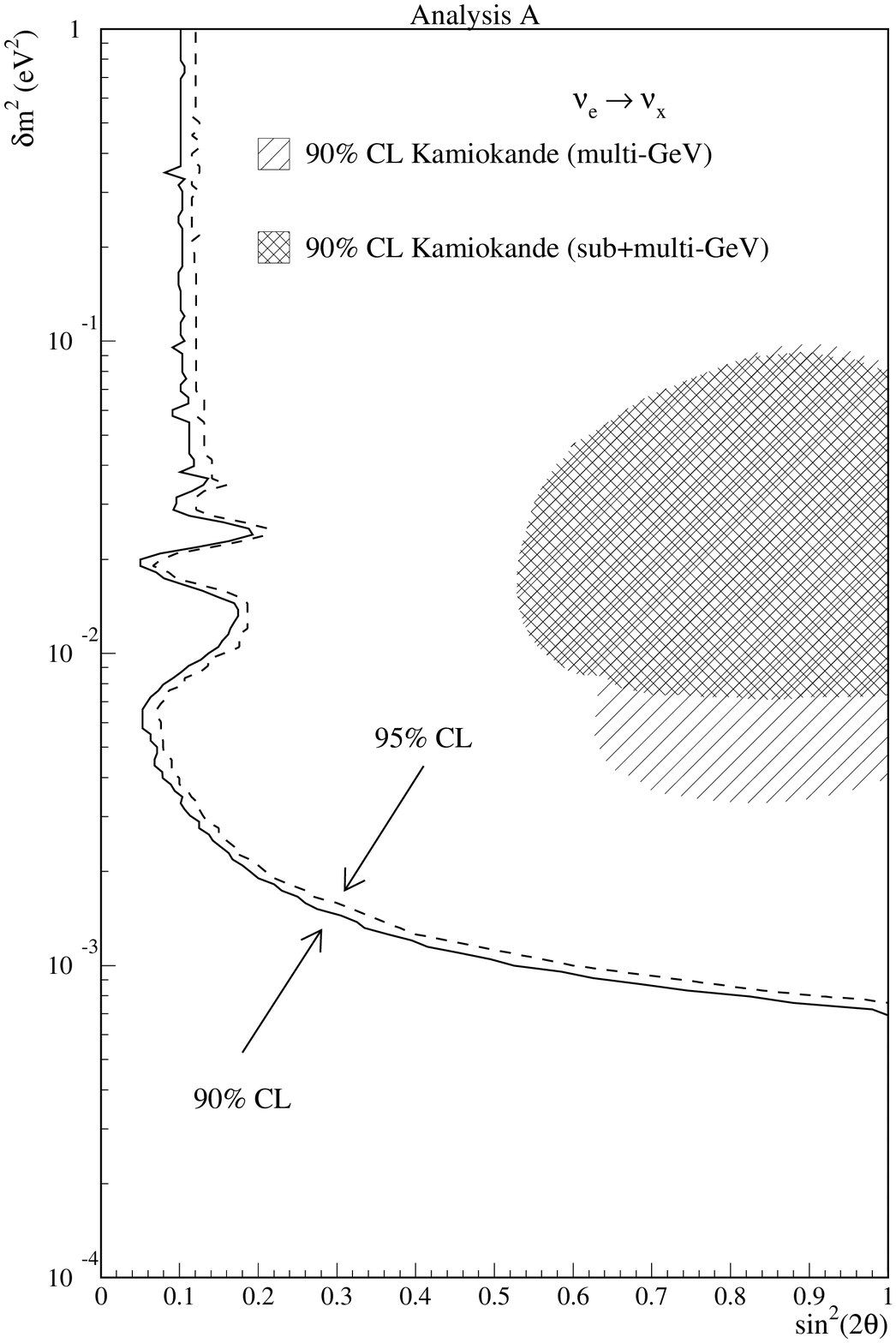} 
\includegraphics[width=6cm,height=8cm]{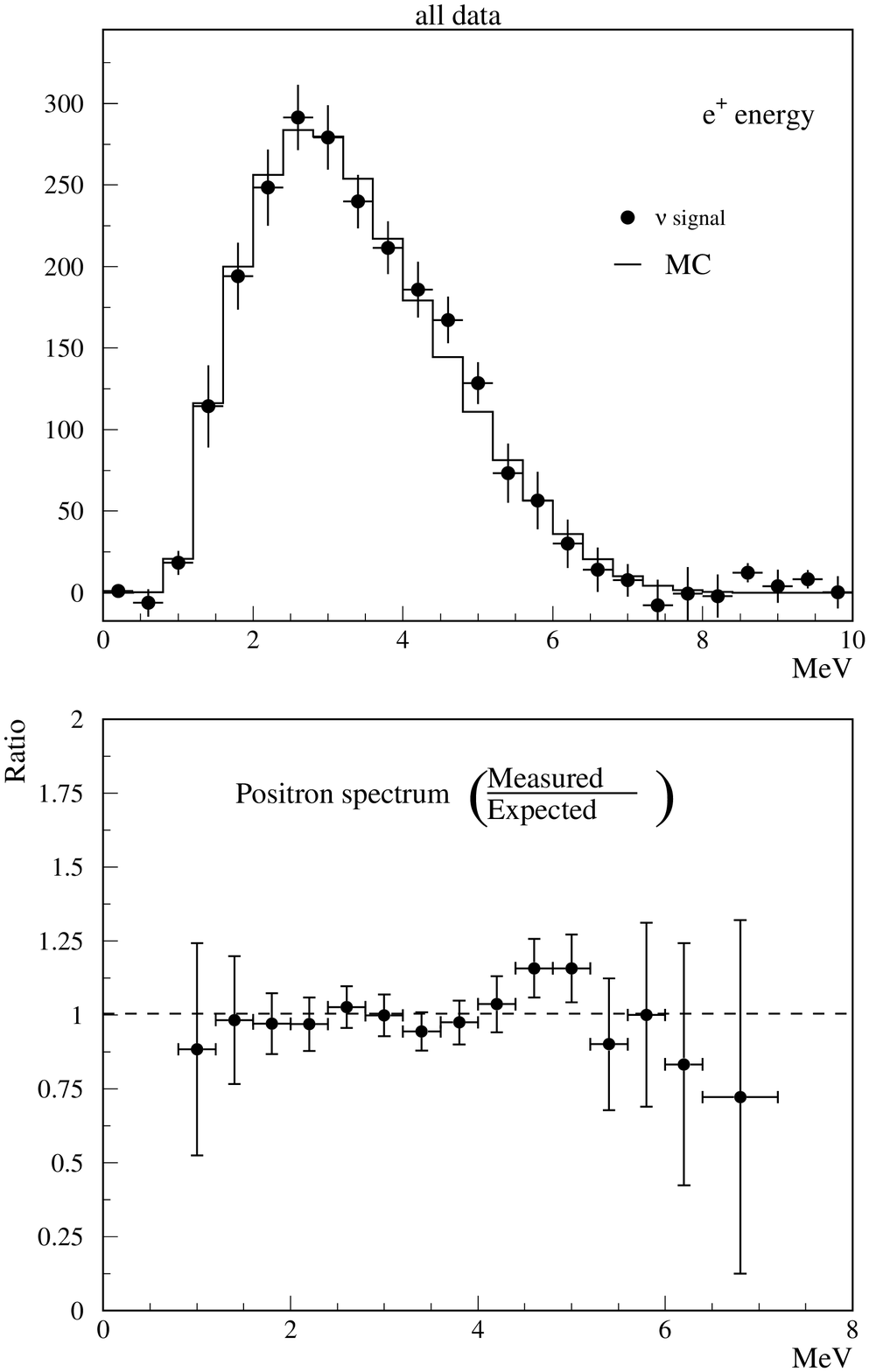}\\
\caption{The CHOOZ exclusion plot in the
  $|\Delta m^2_{\rm 31}| - \sin^22\theta_{13}$ plane (left panel)
  and the measured $e^+$ spectrum normalised to the expected spectrum
   (right panel).
  See \protect\cite{CHOOZ1999} for details.
  }
\label{fig:CHOOZ}
\end{figure}
%

The data accumulated in the neutrino oscillation experiments
by 2003 allowed to determine the parameters
$|\Delta m^2_{\rm 31}|$, $\Delta m^2_{21}$ and
$\sin^2\theta_{12}$ with a relatively
good precision, while $\sin^2\theta_{13}$ was tightly
constrained: $|\Delta m^2_{31}| \cong 2.6\times 10^{-3}~{\rm eV^2}$,
$\Delta m^2_{21}| \cong 7.2\times 10^{-5}~{\rm eV^2}$,
 $\sin^2\theta_{12}\cong 0.3$.
The sign of $\Delta m^2_{\rm 31}$
could not (and still cannot) be determined from the
data and the two possible signs correspond, as it is well known,
to two possible types of neutrino mass spectrum
(see, e.g., \cite{Tanabashi:2018oca}):
with normal ordering (NO), $m_1 < m_2 < m_3$,
and with inverted ordering (IO), $m_3 < m_1 < m_2$.
Depending on the value of the lightest neutrino mass,
the spectrum can be i) normal hierarchical (NH),
$m_1 \ll m_2 < m_3$, ii) inverted hierarchical (H),
$m_3 \ll m_1 < m_2$, or iii) quasi-degenerate (QD),
$m_1 \cong m_2 \cong m_3$ with $m^2_{1,2,3} >> |\Delta m^2_{\rm 31}|$.
The latter requires $m_{1,2,3}\gtap 0.10$ eV.

These developments led to the important realisation that 
the $\betabeta-$decay experiments have a remarkable
physics potential 
\cite{bb0nuPhysPot1,Vissani:1999,Pascoli:2002xq,bb0nuPhysPot2}. 
They can probe the Majorana nature
of massive neutrinos, and if neutrinos are
proven to be Majorana particles, via the measurement 
of  $\meff$ they can provide information on:\\
-- the neutrino mass spectrum (NH, IH, QD),\\
-- absolute neutrino mass scale,\\
-- and with 
input on the value of ${\rm min}(m_j)$ -- on the Majorana phases in 
the PMNS matrix.\\
This best is illustrated by Fig. \ref{fig:meffSP} 
in which $\meff$ is shown as a function 
of ${\rm min}(m_j)$
~\footnote{Figures exhibiting $\meff$ versus ${\rm min}(m_j)$  
appeared first in \cite{Vissani:1999} in a form   
in which (in contrast to Fig. \ref{fig:meffSP})  
only the contours delimiting 
the areas corresponding to NO and IO  spectra 
are shown.}.

In 2006 the quantitative studies performed in \cite{PPRio106},
which were based on advances in leptogenesis theory \cite{Abadaetal},
have shown that the CP violation necessary
for the generation of the observed baryon asymmetry of the 
Universe in the leptogenesis scenario can be provided
exclusively by the Majorana
phases of the neutrino mixing matrix $U$.

  Two experiments with  $^{76}$Ge, IGEX \cite{Aalseth:2002} and
Heidelberg-Moscow \cite{HM:2001}, published significantly
improved limits on $T^{0\nu}_{1/2}(^{76}Ge)$
in 2002 and 2001, respectively.
The IGEX experiment announced first results in 1996. 
In its final configuration it took data in the
Canfranc tunnel, Spain, and used 6.5 kg Ge enriched in $^{76}$Ge at 86\%. 
The following lower limit was obtained \cite{Aalseth:2002}:
$T^{0\nu}_{1/2}(^{76}Ge) > 1.57\times 10^{25}$ yr (90\% C.L.).
A somewhat better limit was reported in 2001 by the Heidelberg--Moscow
collaboration whose detector consisted of 11 kg of Ge
enriched in $^{76}$Ge at 86\% \cite{HM:2001} and was located 
in Gran Sasso Laboratory, Italy:
$T^{0\nu}_{1/2}(^{76}Ge) > 1.9\times 10^{25}$ yr (90\% C.L.).
This result was based on 47.4 kg y of data.

 Analising a somewhat larger data set
(54.98 kg y), H.V. Klapdor-
\begin{figure}
\centerline{\includegraphics[width=0.6\linewidth]{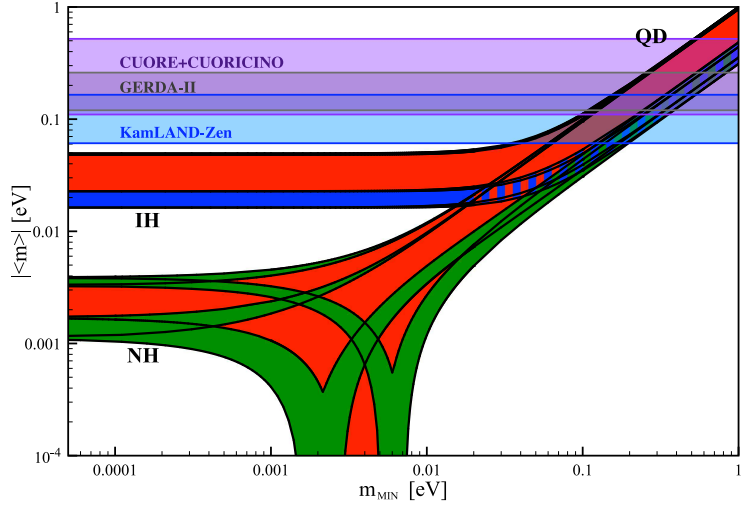}}
\caption{The effective Majorana mass $\meff$
(including a 2$\sigma$ uncertainty),
as a function of ${\rm min}(m_j)$.
The figure is obtained using 
the best fit values and the 1$\sigma$ 
ranges of allowed values 
of  $\Delta m^2_{21}$, $\sin^2\theta_{12}$, 
$\sin^2\theta_{13}$
and $|\Delta m^2_{31(32)}|$ 
from \protect\cite{Capozzi:2017ipn}
(propagated to $\meff$ 
and then taking a 2$\sigma$ uncertainty). 
The phases $\alpha_{21}$ and $(\alpha_{31} - 2\delta)$
are varied in the interval [0,2$\pi$].
The predictions for the NH, IH and QD
spectra  as well as the GERDA-II  \protect\cite{Agostini:2018tnm}, 
KamLAND-Zen \protect\cite{KamLAND-Zen:2016pfg}
and the combined CUORE+CUORICINO \protect\cite{Alduino:2017ehq} limits 
(Eqs. (\ref{eq:GERDA}), (\ref{eq:meffKZen}) and (\ref{eq:meffCCUORE0})),
are indicated. The black lines determine the ranges of
values of $\meff$ for the
 different pairs of CP conserving values of
$\alpha_{21}$ and $(\alpha_{31} - 2\delta)$:
$(0,0)$, $(0,\pi)$, $(\pi,0)$ and  $(\pi,\pi)$.
The red regions correspond to at least one of
the phases $\alpha_{21}$ 
and $(\alpha_{31} - 2\delta)$
having a CP violating value, while the 
blue and green areas correspond to 
$\alpha_{21}$ and  $(\alpha_{31} - 2\delta)$ 
possessing CP conserving values 
(Update by S. Pascoli in \protect\cite{Tanabashi:2018oca}
of a figure from \protect\cite{PPNH07}.)
}
\label{fig:meffSP}
\end{figure}
%
\noindent Kleingrothaus with three members of
the  Heidelberg--Moscow collaboration claimed
the observation of $\betabeta-$decay
first (in 2001) with a half-life \cite{KK:2001}
$T^{0\nu}_{1/2}(^{76}Ge)=1.5\times 10^{25}$ yr,  
later (in 2004 using 71.7 kg y of data) 
with a half-life \cite{KK:2004}
T$^{0\nu}_{1/2}$($^{76}$Ge)$=$1.19$\times 10^{25}$ yr,
and finally in 2006 (reanalising the 2004 data with 
one co-author) with a half-life \cite{KK:2006}
$T^{0\nu}_{1/2}(^{76}Ge)=2.23_{-0.31}^{+0.44}\times 10^{25}$ yr.
The Moscow part of the  Heidelberg--Moscow collaboration
disagreed with this claim \cite{Bakalyarov:2005}.

 The analyses leading to the claims of observation of $\betabeta-$decay by
Klapdor-Kleingrothaus et al. were criticised in a number of articles
(see  \cite{MGoodman} for further details). 
The claimed positive results on  $\betabeta-$decay published by
Klapdor-Kleingrothaus et al. in \cite{KK:2001,KK:2004,KK:2006}
were definitely proven to be incorrect by the results of GERDA II experiment
with  $^{76}$Ge \cite{Agostini:2018tnm}:
\begin{eqnarray} 
{\rm T^{0\nu}_{1/2}(^{76}Ge)} > 8.0\times 10^{25}~{\rm yr}~(90\%~{\rm C.L})\,,
\\[0.20cm]
\meff < (0.16 - 0.26)~{\rm eV}\,,~~{\rm GERDA~II}\,,
\label{eq:GERDA}
\end{eqnarray}
%
where we have quoted also the limit on the effective Majorana mass
$\meff$ reported by the GERDA II collaboration. The interval reflects
the estimated uncertainty in the relevant NME used to extract
the limit on $\meff$.

Two experiments CUORICINO (using cryogenic detector with bolometers)
and NEMO-3 (employing tracking device) took data in the discussed
period and were completed respectively 
\begin{figure}
\centerline{\includegraphics[width=0.8\linewidth]{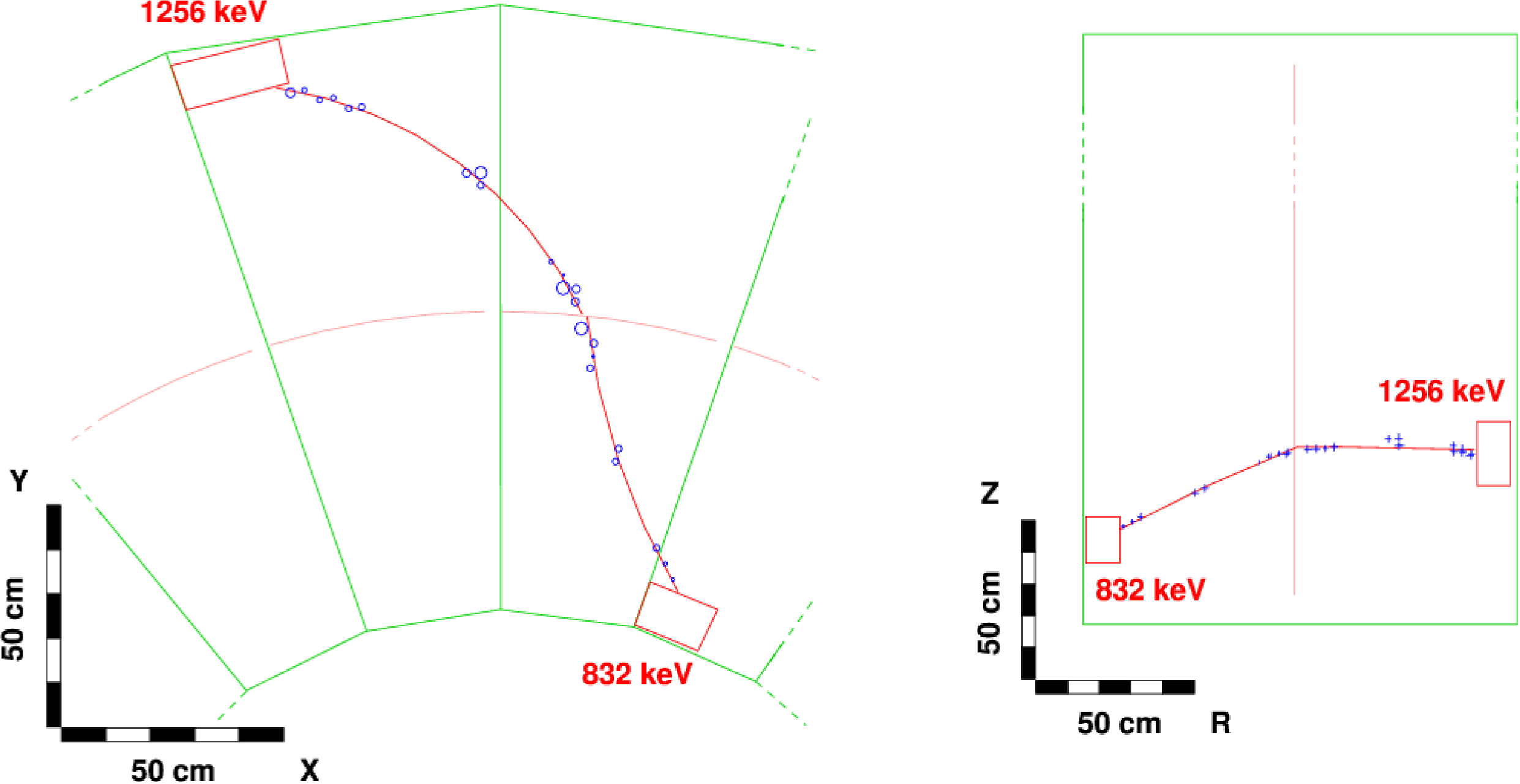}} 
\caption{A $\betabetanu-$decay event as seen by the NEMO-3 detecor.}
\label{fig:NEMO3}
\end{figure}
%
\noindent in 2008 and 2011.
The CUORICINO collaboration used 40.7 kg of  $^{130}$Te 
and obtained the following  final result \cite{Arnaboldi:2005}:
$T^{0\nu}_{1/2}(^{130}Te) > 2.8 \times 10^{24}$ yr.

 The NEMO-3 experiment searched for $\betabeta$-decay and
observed the $\betabetanu-$decay of 
$^{48}$Ca, $^{82}$Se, $^{96}$Zr, $^{100}$Mo, $^{116}$Cd, $^{130}$Te and 
$^{15}$Nd \cite{Arnold:2006}. The best limit was on
$\betabeta$-decay half-life was obtained with $^{100}$Mo
(6.914 kg source):
$T^{0\nu}_{1/2}(^{100}Mo) > 1.1\times 10^{24}$ yr (90\% C.L.).
In this experiment more than 700 000 (!) $\betabetanu-$decays
of $^{100}$Mo were detected with almost zero background
and the  individual energy spectra and angular 
distribution of electrons were measured 
\footnote{The images of $\betabetanu-$decay from NEMO-3 
detector have certain delicate elegance, 
see Fig. \ref{fig:NEMO3}.}. 
The $\betabetanu$-decay half-lives of 
$^{48}$Ca, $^{82}$Se, $^{96}$Zr, $^{100}$Mo, $^{116}$Cd, $^{130}$Te and 
$^{15}$Nd were also determined with impressive precision
\footnote{
These results are published in a large number of publications
the last of which appeared in 2018. In the present article only 
four are quoted (see \cite{Arnold:2006}).
}.

Values of  $(\beta\beta)_{2\nu}$ half-life $T^{2\nu}_{1/2}$ of different nuclei 
measured by 2010 are shown in Table \ref{tab:Tbb2nu} (taken from 
\cite{Barabash:2010}).

%
\section{Latest Results, Current Problems and Prospects}   
%

 We have quoted the latest result from the Gerda II experiment 
in Eq. (\ref{eq:GERDA}), which represents the best 
limit on $\betabeta-$decay half-life of  $^{76}$Ge.
We quote next two more best limits 
on $\betabeta-$decay half-lives of $^{130}$Te  and 
$^{13e}$Xe, 
obtained respectively in the combined analysis of CUORICINO and CUORE-0 
data \cite{Alduino:2017ehq} and in the KamLAND-Zen experiment 
\cite{KamLAND-Zen:2016pfg}:
\begin{eqnarray}
\nonumber
& {\rm T^{0\nu}_{1/2}(^{130}Te)} > 1.5\times 10^{25}~{\rm yr}~(90\%~{\rm C.L})\,,
\\[0.20cm]
& \meff < (0.11 - 0.52)~{\rm eV}\,,~~{\rm CUORICINO+CUORE-0}\,;
\label{eq:meffCCUORE0}
\end{eqnarray}
\begin{eqnarray}
\nonumber
& {\rm T^{0\nu}_{1/2}(^{136}Xe)} > 1.07\times 10^{26}~{\rm yr}~(90\%~{\rm C.L})\,,
\\[0.20cm]
& \meff < (0.061 - 0.165)~{\rm eV}\,,~~{\rm KamLAND-Zen}\,,
\label{eq:meffKZen}
\end{eqnarray}
%
The intervals reflect the
estimated uncertainties in the relevant 
NMEs used to extract the limits on 
$\meff$ from the experimentally obtained lower
bounds on the $^{130}{\rm Te}$ and  $^{136}{\rm Xe}$ 
$\betabeta-$decay half-lives
\footnote{For a review of the limits on 
$\meff$ obtained in other $\betabeta$-decay 
experiments and a detailed discussion of the 
NME calculations for  $\betabeta$-decay  and
their uncertainties see, e.g., 
\cite{Vergados:2016hso}.}.

Important role in the searches for $\betabeta-$decay 
in the discussed period 
was played by the EXO experiment with liquid 
$^{136}{\rm Xe}$ time projection chamber \cite{EXO:2018}.
In its final phase (EXO-200), the data was obtained 
with $\sim$100 kg of $^{136}{\rm Xe}$.
The following best limit was reported by EXO 
collaboration \cite{Albert:2017owj}: 
$T^{0\nu}_{1/2}(^{136}Xe) > 1.8\times 10^{25}$ yr (90\% C.L).

 The ``conservative'' upper limit $\meff = 0.165$ eV 
(Eq. (\ref{eq:meffKZen})),
which is in the range of the QD neutrino mass spectrum,  
implies the following upper limit on the absolute 
Majorana neutrino mass scale \cite{Penedo:2018kpc}:
$ m_0 \cong m_{1,2,3} \ltap 0.60$ eV.

  A large number of experiments of a new generation aims at a 
sensitivity to  $\meff \sim (0.01 - 0.05)$~eV
\footnote{An incomplete list includes 
CUORE ($^{130}$Te),
GERDA II ($^{76}$Ge), MAJORANA ($^{76}$Ge), 
LEGEND ($^{76}$Ge),
KamLAND-ZEN ($^{136}$Xe),
nEXO ($^{136}$Xe), 
SNO+ ($^{130}$Te),
AMoRE ($^{100}$Mo),
CANDLES ($^{48}$Ca),
SuperNEMO ($^{82}$Se, $^{150}$Nd),
NEXT ($^{136}$Xe),
DCBA ($^{82}$Se, $^{150}$Nd),
PANDAX-III ($^{136}$Xe),
ZICOS ($^{96}$Zr),
MOON ($^{100}$Mo).},
which will allow to probe the whole
range of the predictions for  $\meff$  in the case of 
IO neutrino mass spectrum \cite{Pascoli:2002xq} shown 
in Fig. \ref{fig:meffSP} (for reviews of the currently 
running and future planned
$\betabeta$-decay  experiments and their prospective
sensitivities see, e.g., 
\cite{Vergados:2016hso,Abgrall:2017syy,Dolinski:2019nrj}).
\begin{table}[t]
\caption{Values of $T^{2\nu}_{1/2}$ measured by 2010. 
(Table taken from \protect\cite{Barabash:2010} 
where one can find details of how the values 
quoted in the table were obtained.)}
\label{tab:Tbb2nu}
\vspace{0.4cm}
\begin{center}
\begin{tabular}{l|r}
\hline
\multicolumn{1}{c|}{Isotope} &\multicolumn{1}{c}{$T^{2\nu}_{1/2}$ y} \\
\hline
$^{48}$Ca &$4.4^{ + 0.6} _{-0.5 }\times 10^{19}$ \\
$^{76}$Ge&$(1.5\pm 0.1) \times 10^{21}$ \\
$^{82}$Se&$(0.92 \pm 0.07)\times 10^{20}$ \\
$^{96}$Zr&$(2.3\pm 0.2) \times 10^{19 }$ \\
$^{100}$Mo&$(7.1 \pm 0.4)\times 10^{18}$ \\
$^{100}$Mo--$^{100}$Ru(0$^{ + }_{1}$) &$5.9^{ + 0.8}_{ - 0.6} \times 10^{20 }$ \\
$^{116}$Cd&$(2.8 \pm 0.2)\times 10^{19}$ \\
$^{128}$Te&$(1.9 \pm 0.4)\times 10^{24}$ \\
$^{130}$Te&$6.8^{ + 1.2}_{-1.1}\times 10^{20}$ \\
$^{150}$Nd&$(8.2 \pm 0.9)\times 10^{18 }$ \\
$^{150}$Nd--$^{150}$Sm(0$^{ + }_{1}$)&$1.33^{ + 0.45}_{ - 0.26 }\times 10^{20}$ \\
$^{238}$U&$(2.0 \pm 0.6)\times 10^{21}$ \\
$^{130}$Ba, ECEC(2$\nu $)&$(2.2 \pm 0.5)\times 10^{21}$ \\
\hline
\end{tabular}
\end{center}
\end{table}
%

 Obtaining quantitative information on
the neutrino mixing parameters from a measurement of
$\betabeta$-decay half-life would be impossible 
without sufficiently precise knowledge of the 
corresponding NME of the process.
At present the variation of the values
of different $\betabeta$-decay  NMEs 
calculated using various currently employed methods 
is typically by factors $\sim (2-3)$ 
(for a discussion of the current status of the 
calculations of the NMEs for the 
$\betabeta$-decay see, e.g., 
\cite{Vergados:2016hso,Dolinski:2019nrj,Engel:2016xgb}).

 Additional source of uncertainty 
is the effective value of the axial-vector 
coupling constant $g_A$ in $\betabeta-$decay. 
This constant is related to the weak charged axial current 
(Gamow-Teller transitions), which is not conserved and therefore 
can be and is renormalised, i.e., quenched,
by the nuclear medium. This implies that $g_A$ is reduced 
from its free value of $g_A = 1.269$.
The reduction of $g_A$ can have important 
implications for the $\betabeta-$decay searches since 
to a good approximation  $T^{0\nu}_{1/2} \propto (g^{eff}_A)^{-4}$.
The problem of the $g_A$ quenching arose in connection with the 
efforts to describe theoretically the experimental 
data on $\betabetanu-$decay \cite{Iachello:2015ejm}.
The physical origin of the quenching is not fully understood, 
and the magnitude of the quenching of $g_A$ in 
$\betabeta-$decay is subject to debates 
(for further details see, e.g., \cite{Vergados:2016hso}).

  The  $\betabeta-$decay can be generated, in principle, 
by a $\Delta L = 2$ mechanism other than the light Majorana neutrino 
exchange considered here, or by a combination of mechanisms one of 
which is the light Majorana neutrino exchange (for a discussion 
of different mechanisms which can trigger $(\beta\beta)_{0\nu}$-decay, 
see, e.g., \cite{bb0nuOther,bb0nuMultiple1} 
and the articles quoted therein). 
Actually, the predictions for 
$\meff$  in the cases of the NH, IH and QD
neutrino mass spectra (shown in Fig. \ref{fig:meffSP})
can be drastically modified  
by the existence of lepton charge non-conserving ($|\Delta L| = 2$)
new physics beyond that predicted by the SM: eV or GeV to TeV scale 
RH Majorana  neutrinos, etc. 
(see, e.g., \cite{Bilenky:2001xq}). 
There is a potential synergy between the searches for 
$\betabeta-$decay and the searches for neutrino-related 
$|\Delta L| = 2$ beyond the SM physics at LHC:
$\betabeta-$decay experiments with a sensitivity
to half-lives of $T^{0\nu}_{1/2} = 10^{25}$ yr probe approximately 
values of $\meff \sim 0.1$ eV and ``new physics''  
at the scale $\Lambda_{LNV}\sim 1$ TeV
(see, e.g., \cite{Helo:2013ika} and references quoted therein).

\section{Epilogue}

The authors of the first experiment searching 
for $\betabeta$ and $\betabetanu$ decays in the USSR 
\footnote{The experiment was performed with $^{48}$Ca.}, 
Dobrokhotov et al., wrote in 1956 \cite{Dobrokhotov:1956}:
``The search for double beta-decay is an amazing example of a 
fantastic succession of periods of hope and of disillusionment. 
Two times in the course of a single decade this phenomenon has 
been ``discovered'', and both times the discovery
has been found to be erroneous. The history of the question is still not
complete; the phenomenon has not been observed experimentally, 
and the succession of journal articles in recent years only gradually 
sets larger and larger lower limits on the lifetime of a nucleus 
capable of double beta-decay. In the present
research we have again not succeeded in observing the event, 
but the limit of the half-life of the process  
has been raised to about $0.7\times 10^{19}$ years, and further
steps in this direction (if indeed they are worth-while) 
will depend on achieving considerable increases in the amount 
of material subjected to study.''

\vspace{0.2cm}
 The situation regarding the experimental searches for 
$\betabeta-$decay is essentially the same today.
 The quest for one of the deepest ''secrets'' of Nature 
-- the nature (Dirac or Majorana) of massive neutrinos and for the status of 
lepton charge conservation -- continues. 
%
\section*{Acknowledgements}
%
The author would like to thank M. Cribier, J. Dumarchez,
D. Vignaud and their colleagues for the organisation of the timely
and very interesting conference on ``History of the Neutrino''. 
Useful discussions with F. \v Simkovic of the NME calculations
for $\betabeta$ and $\betabetanu$ decays 
are acknowledged with gratefulness.
This work was supported in part by the 
European Union's Horizon 2020 
research and innovation programme under the Marie 
Sklodowska-Curie grant agreements No 674896 
(ITN Elusives) and No 690575 (RISE InvisiblesPlus), 
by the INFN program on Theoretical Astroparticle Physics 
and by the  World Premier International Research Center
Initiative (WPI Initiative, MEXT), Japan.

\end{document}